\documentclass[preprint,aps]{revtex4}

\usepackage{graphicx}
\begin{document}

\title{Bulk phantom fields, increasing warp factors and fermion localisation}
\author{
Ratna Koley \footnote{Electronic address: {\em ratna@cts.iitkgp.ernet.in}}
${}^{}$ and Sayan Kar \footnote{Electronic address : {\em sayan@cts.iitkgp.ernet.in}
}${}^{}$}
\affiliation{Department of Physics and Centre for
Theoretical Studies \\Indian Institute of Technology, Kharagpur 721 302, India}

\begin{abstract}

A bulk phantom scalar field (with negative kinetic energy) in a
sine--Gordon type potential is
used to generate an exact thick brane solution with an
increasing warp factor. It is shown that the growing nature of the
warp factor allows the localisation of massive as well as massless
spin-half fermions on the
brane even without any additional non--gravitational interactions.
The exact solutions for the localised massive fermionic modes are
presented and discussed. The inclusion of a fermion--scalar Yukawa
coupling appears to change the mass spectrum and wave functions 
of the localised fermion though it does not play the crucial role it did
in the case of a decreasing warp factor.

\end{abstract}

\maketitle

An important issue in the study of braneworld models {\cite{rs}
is the question of localization of the Standard Model fields on the brane. 
This problem was addressed, many years ago, 
by Rubakov and Shaposhnikov and  independently by
Akama {\cite{ruba}} though the effect of higher dimensional gravity
was not included there. With the advent of Randall--Sundrum (RS)
models where a warped higher dimensional background (bulk) geometry
is assumed, there has been 
renewed interest along these directions. The extra dimension in the
RS scenario can be finite (RS-I) or infinite (RS-II). We shall mainly
deal with RS-II type models in this article. 
 
Our first goal here is to generate a thick brane realisation of RS-II with
a growing warp factor. Such increasing warp factor
solutions have been obtained by several authors {\cite{increasing}}. Here
we propose a hypothetical
scalar field with a wrong-sign (negative) kinetic energy in the 
bulk. Such strange `matter' fields have been named as 
the `phantom' or `ghost' and has received increasing attention 
among theorists because of their potential in explaining 
the observational accelerated expansion of our universe. 
It has been introduced by 
Caldwell {\cite{cald}} and others {\cite{carr}} in the context of four dimensional
cosmology. The phantom field obviously violates all versions
of the energy conditions--strong, weak, null as well as dominant.
Given our limitation in obtaining  
any experimental evidence for bulk matter one may
propose a model in which the bulk consists of a energy condition 
violating scalar field. If the four dimensional consequences of
such exoticity in the fifth seem reasonable, we should perhaps
not worry too much about what is really there in the fifth dimension. 
Having found the warp factor, we then look at the localisation of
fermions. It turns out that the fermion spectrum can be exactly obtained
and there do exist localised states. 
Furthermore, we show that massless as well as
massive modes of the spin 1/2 fermion can be confined on the 3-brane
even without any kind of Yukawa
interaction between the bulk and brane matter fields.

Let us begin with
the action for the bulk spacetime which consists of a negative
cosmological constant, gravity and a bulk scalar field with a
sine-Gordon potential:  

\begin{equation}
S = \int \left [ \frac{1}{2\kappa_5^2}\left (R-2\Lambda\right ) + \frac{1}{2}
(\nabla \phi)^2 - V(\phi) \right ] \sqrt{-g} d^5 x
\end{equation}
  
Note that the kinetic energy of the bulk scalar field is set to be
negative. 

Our setup has a single brane embedded in the bulk.
The ansatze for a warped  
spacetime following standard terminology, is given by:

\begin{equation}
ds^2 = d\sigma^2 +e^{-2f(\sigma)} \eta_{ij}dx^i dx^j
\end{equation}

where, the signature convention is (- + + + +) and the small Latin
indices $i, j = 0, 1, 2, 3$
refer to the brane coordinates. The scalar field will be assumed to be
a function of the extra coordinates only. 

In this framework, the Einstein scalar system reduces to the following set of coupled, 
nonlinear ordinary differential equations :

\begin{eqnarray}
f '' & = & - a {\phi'}^2 \\
{f'}^2 & = & \frac{a}{4} \left ( - {\phi'}^2 - 2V \right ) -\frac{\Lambda}{6}\\
\phi'' -4f'\phi' &  = & - \frac{dV}{d\phi}
\end{eqnarray}

where $a=\kappa_5^2/3 $.

The scalar field (Klein-Gordon) equation follows from the first two Einstein
equations and is not independent. 
This system of equations can be solved exactly for a sine-Gordon potential 
given by 

\begin{equation}
V(\phi) = B \left (1 + \cos\frac{2 \phi}{A} \right)
\end{equation}

The explicit forms of the solutions for the scalar field and the warp factor are given
as  

\begin{eqnarray}
f (\sigma) = -  \frac{a}{\kappa_1} \sqrt{\frac{\vert\Lambda \vert}{6}}\ln
\cosh \left (\frac{\kappa_1}{a}\sigma\right ) \\
\phi(\sigma) = 2A \tan^{-1}\left ( \exp{\frac{\kappa_1}{a}\sigma} \right )
-\frac{\pi A}{2}
\end{eqnarray}

where, $\kappa_1 = \frac{1}{A^2}\sqrt{\frac{\vert \Lambda \vert}{6}}$. 
And the constant B in the sine-Gordon potential is given as :
$B= \frac{\vert \Lambda\vert}{6a^2} \left (a-\frac{1}{4A^2}\right )$. 
Note here that a decreasing warp factor is obtainable through a usual
(positive kinetic energy) scalar field in a SG potential. This is discussed
by us in {\cite{rksk}}. 

It is important to note the presence of a negative cosmological constant
in both the warp factor and the soliton.
This model with a bulk SG potential provides a `thick brane' 
scenario where the SG field and its soliton
configuration dynamically generate this domain wall configuration
in the background warped geometry. In addition as is obvious from the
functional from of $f$, there is no discontinuity in the derivative of
$f$ at the location of the brane. The warp factor is smooth everywhere
and has all the necessary features.    
 
The line element in the warped bulk space time is given by

\begin{equation}
ds^{2} = d\sigma^{2} + \cosh^{2\nu} \left (\frac{\kappa_1}{a}\sigma
\right )  \eta_{ij} dx^i dx^j
\end{equation}

where $\nu = \frac{a}{\kappa_1} \sqrt{\frac{\vert\Lambda \vert}{6}}={A^2}a$. 
The distinct feature of this metric is that the warp factor is
a growing function of the extra coordinate. The 
phantom field is responsible for such a line element for the 
background geometry. As we show later, 
the localisation of spinor fields 
on the brane without any non--gravitational interaction 
with the bulk scalar is achieved largely due to the growing nature of the
warp factor.  

Notice that the metric is completely non-singular for the full domain of
the fifth coordinate. It describes a space of negative Ricci curvature
where $R$ is given by :

\begin{equation}
R = - 4 \frac{\vert \Lambda \vert}{6} \left[ \left (5 - \frac{2}{\nu} \right ) 
\tanh^2 \left (\frac{\kappa_{1}}{a} \sigma \right ) + \frac{2}{\nu} \right]
\end{equation}

It is straightforward to check that R is always negative and  
asymptotically ($\sigma \rightarrow \pm \infty$) takes on the
value of $-\frac{10}{3}\vert \Lambda \vert$. One can also get an AdS space 
of {\em constant} negative Ricci curvature for the value of the parameter 
$\nu = 2/5$. 


\begin{figure}[htb]
\includegraphics[width= 10cm,height=5.5cm]{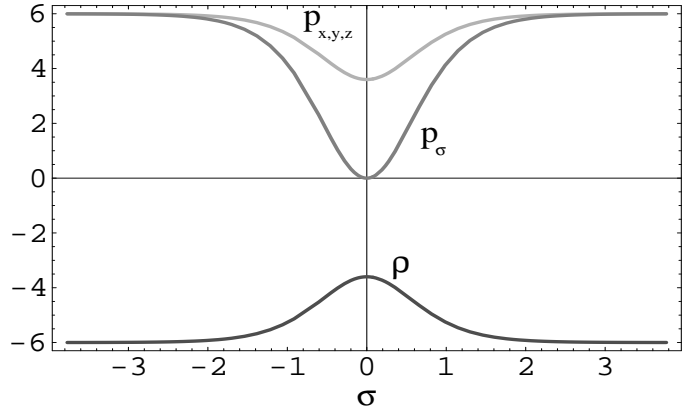}
\caption{Variation of the energy density  $\rho$ and pressure ( $p_i$ and 
 $p_\sigma$ ) with $\sigma$ corresponding to the warped geometry given by the 
metric in eqn. $ (9) $ for $ \frac{\kappa_{1}}{a} = \frac{4}{5}$ and $\vert \Lambda \vert =$ 6.} 
\end{figure}

The energy density and pressures, which generate such a line
element, are found to be :

\begin{equation}
\rho = - p_{x,y,z} = - 3 \frac{\kappa_1}{a} \sqrt{\frac{\vert \Lambda \vert}{6}}
\mbox{sech}^2\frac{\kappa_1}{a}\sigma - \vert \Lambda \vert \tanh^2\frac{\kappa_1}{a}
\sigma
\end{equation}
\begin{equation}
p_{\sigma}= \vert \Lambda \vert \tanh^2\frac{\kappa_1}{a}\sigma
\end{equation}

At $\sigma=0$ (i.e. the location of the brane)
we have $p_{x,y,z}=-\rho$ and $p_{\sigma}=0$ -- an effective cosmological
constant on the 3--brane. The energy density and pressures are plotted in
Figure (1). It is worth noting that the matter stress-energy which acts
as a source for the warped geometry does not satisfy the Null Energy Condition
(NEC $\rightarrow$  $ \rho+p_i\ge 0$). The other energy conditions
(such as the Weak Energy Condition and the Dominant Energy Condition) are
 also viotated. 
{\cite{visser}}. 

Let us now focus attention on the localization of spinor fields on the brane. 
It is known that the kink 
solutions of the scalar field as given in may trap fermionic zero 
modes {\cite{rebbi}}. 
The bulk fermion coupled to the scalar field in 5D gives rise to two chiral fermionic zero modes 
in the four dimensions. Depending on the sign of the coupling constant one of these two modes 
is found to be localized on the brane while the other is delocalized and not normalizable 
\cite{ringeval}. In all the models discussed till date,
the Yukawa coupling with the bulk scalar is capable of 
localising only a single chiral state (right or left) while
fermions of both chiralities would be expected. We will show that in the 
geometry with an increasing warp factor both the left and right chiral 
zero modes are confined to the brane even
without any coupling with the bulk field. 
Localised massive modes are also achieved in this background geometry. 

The Lagrangian for a Dirac fermion propagating in the five dimensional warped 
space with the metric (9) is :

\begin{equation}
\sqrt{-g} {\cal{L}}_{Dirac} = \sqrt{-g}\hspace{.03in}(i \bar{ \Psi} \Gamma^{a} {\cal{D}}_{a} \Psi
- \eta_{F} \bar{ \Psi} \mbox{F}(\Phi) \Psi )
\end{equation}

where $g = \mbox{det}(g_{ab})$, is the determinant of full five dimensional metric and $\Phi=\phi/A$. 
The matrices $\Gamma^{a} = (e^{f(\sigma)} \gamma^{\mu}, -i \gamma^{5})$ provide a 
four dimensional
representation of the Dirac matrices in five dimensional curved space. Where $\gamma^{\mu}$
and $\gamma^{5}$ are the usual four dimensional Dirac matrices in chiral representation.
We also consider a coupling between the bulk
scalar and Dirac field. Later we will show that, it is not
necessarilly required for the localization of fermions on the brane. 
However, we will discuss the effect of the coupling on the confinement of
different fermionic modes on the brane. 

The covariant derivative in 5D curved space for the metric given
in Eqn. (9) {\cite {wb}}:

\begin{equation}
{\cal{D}}_{\mu} =(\partial_{\mu} -\frac{1}{2} f'(\sigma) e^{-f(\sigma)} \Gamma_{\mu}
\Gamma^{4}) ; \hspace{.8cm}
{\cal{D}}_{4} = \partial_{\sigma}
\end{equation}

The Dirac Lagrangian in 5D curved spacetime then reduces to the following form
 
\begin{equation}
\sqrt{-g} {\cal{L}}_{Dirac} = e^{- 4 f (\sigma)}  \bar{\Psi}\left[i e^{f(\sigma)}\gamma^{\mu}\partial_{\mu} + 
\gamma^{5} (\partial_{4} -2 f'(\sigma)) - \eta_{F} \mbox{F}(\Phi) \right ] 
\Psi  
\end{equation}

The dimensional reduction from 5D to 4D is performed in such a way that the
standard four dimensional chiral particle theory is recovered. 
The five dimensional spinor can be decomposed into four dimensional and 
fifth dimensional parts: $\Psi(x^{\mu},\sigma) =
\Psi(x^{\mu}) \xi(\sigma)$. Since the four dimensional massive
fermions require both the left and right chiralities
it is convenient to organise the spinors with respect to
$\Psi_{L}$ and $\Psi_{R}$ which represent four component spinors living in five
dimensions given by $\Psi_{L,R} = \frac{1}{2} (1 \mp \gamma_{5})\Psi$.
Hence the full 5D spinor can be split up in the following way

\begin{equation}
\Psi(x^{\mu},\sigma) = \left( \Psi_{L}(x^{\mu})\xi_{L}(\sigma) +
\Psi_{R}(x^{\mu})\xi_{R}(\sigma) \right)
\end{equation}

where $\xi_{L.R}(\sigma)$
satisfy the following eigenvalue equations

\begin{eqnarray}
 e^{-f(\sigma)}\left [\partial_{\sigma}-2 f'(\sigma) - \eta_{F} \mbox{F}(\Phi) \right ] \xi_{R}(\sigma)
& = & -m \xi_{L}(\sigma) \\
 e^{-f(\sigma)} \left [\partial_{\sigma}-2 f'(\sigma) + \eta_{F} \mbox{F}(\Phi) \right ] \xi_{L}(\sigma)
& = & m \xi_{R}(\sigma)
\end{eqnarray}

Here $m$ denotes the mass of the four dimensional fermions.
The full 5D action reduces to the standard four dimensional action 
for the massive chiral fermions, 
when integrated over the extra dimension \cite{ringeval}, if
(a) the above equations are satisfied by the bulk 
fermions and (b) the following orthonormality conditions are obeyed.

\begin{eqnarray}
\int_{-\infty}^{\infty} e^{-3 f(\sigma)} \xi_{L_{m}} \xi_{L_{n}} d\sigma =
\int_{-\infty}^{\infty} e^{-3 f(\sigma)} \xi_{R_{m}} \xi_{R_{n}} d\sigma = \delta_{m n} \\
\int_{-\infty}^{\infty} e^{-3 f(\sigma)} \xi_{L_{m}} \xi_{R_{n}} d\sigma = 0
\end{eqnarray}

The dynamical features of the model 
can thus be obtained from the solutions of the eigenvalue equations 
(17) and (18).     

Let us first focus on massless (i.e. $m = 0$) fermions for a kink--fermion Yukawa coupling of the form
$F(\Phi) = \sin \Phi$. The choice of $F(\Phi)$ is governed by (i) the 
requirement of $\sigma \rightarrow -\sigma$ symmetry of the effective potential (see Eqn. (23) below)
and (ii)
exact integrability of the Schrodinger--like equation. One may also consider $F(\Phi)=\Phi$
or any other odd function of $\Phi$ but in such cases the equation for the fermion wave function
becomes too complicated to integrate analytically.
Equation (17) and (18)
reduce to two decoupled equations. The asymptotic solutions are

\begin{eqnarray}
\xi_{L}(\sigma) & \sim & e^{-\left (\eta_{F}  + 2 \sqrt{\frac{\vert \Lambda \vert}{6}}
\right ) \vert \sigma \vert} \\
\xi_{R}(\sigma) & \sim & e^{ \left (\eta_{F}  - 2 \sqrt{\frac{\vert \Lambda \vert}{6}}
\right ) \vert \sigma \vert}
\end{eqnarray}

It is clear from the above expressions that both the left and right chiral
massless modes are normalizable even for 
$\eta_{F} = 0$, This is in accord with the statements on exponentially
rising warp factors in {\cite{bajc}}.   
Obviously, restrictions on nongravitational couplings are required in the RS
framework (decreasing warp factor) for 
the fermions to be confined on the brane {\cite{daemi,ringeval}}. 

We now turn our attention towards the massive fermions. Using the rescaling
$\tilde\xi_{L.R}(\sigma) = e^{-\frac{3}{2} f(\sigma)} \xi_{L.R} (\sigma)$
and Eqn. (17) and (18) we obtain
a second order decoupled equation for the 
left chiral fermions whereas the right chiral modes can be
completely defined from prior knowledge of the left chiral states.

The equation for $\tilde\xi_{L}(\sigma)$ can be recast as a generalised Schr\"{o}dinger equation
by the suitable scaling of the function 
$\tilde\xi_{L} \rightarrow \hat\xi_{L}(\sigma) e^{f(\sigma)}$, given as follows 

\begin{equation}
\partial^{2}_{\sigma} \hat\xi_{L}(\sigma) + \left[m^{2}e^{2f(\sigma)} +
\frac{f''(\sigma)}{2} + \eta_{F} \frac{\mbox{dF}}{\mbox{d}\Phi}
\Phi'(\sigma)- \left (\eta_{F}\mbox{F}(\Phi) + \frac{f'(\sigma)}
{2} \right )^{2}\right ]
\hat\xi_{L}(\sigma) = 0
\end{equation}

The bulk spacetime is $Z_{2}$ symmetric with respect to the brane at $\sigma$ = 0. 
Consider the Yukawa coupling, F$(\Phi) = \sin \Phi$ mentioned earlier. The 
effective potential acting on the left chiral fermions then shows a minimum at 
the location of the brane and asymptotically approaches a constant positive value. 
This functional form is known as
the P\"{o}sch Teller potential {\cite{Landau}} well 
In the case at hand, this can be written explicitly as : 

\begin{equation}
V_{eff} (\sigma) =  - \left (\eta_{F}^2 + m^2 - \frac{b^2}{4} \right ) \mbox{sech}^2(b \sigma) 
\end{equation}

\begin{figure}[htb]
\includegraphics[width= 10cm,height=5.5cm]{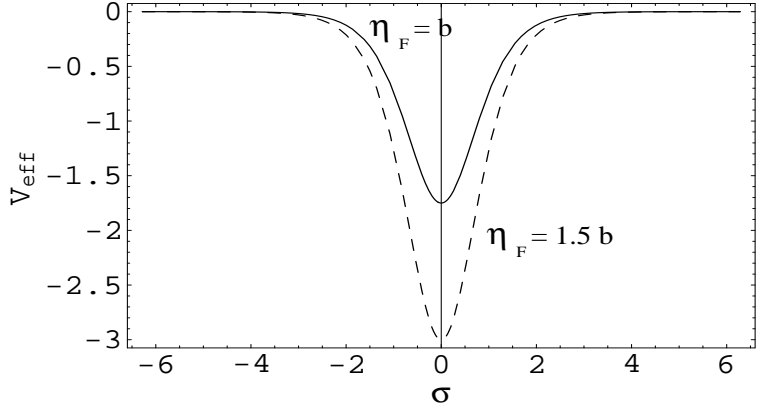}
\caption{The effective potential acting on the left chiral fermions 
plotted as a function of extra dimension $\sigma$ for two different values of 
coupling constant, $\eta_{F} =$ b and 1.5 b which depicts that the depth 
is greater for stronger coupling. In this case we choose $\vert \Lambda \vert = 6$ 
and mass $m =$ b.} 
\end{figure}

where, $b = \frac{\kappa_{1}}{a}$. The nature of the potential is shown in Fig. (2). 
The depth of the potential increases with increasing coupling constant. 
It is evident that the modes will be found to be localised on the brane even for 
$\eta_{F} = 0$. We choose to work with the parameter $\nu = 1$
in our study of the massive modes. This choice is prompted by
integrability of the equation, though, with growing warp factors with
$\nu\neq 1$ the nature of the potential remains unchanged and our
qualitative conclusions will also not be altered.
 
The Eqn. (23) reduces to the following form 

\begin{equation}
\partial^{2}_{{\sigma}} \hat{\xi_{L}}({\sigma}) + 
(A_{1} \mbox{sech}^{2} (b \sigma)  - A_{2})  \hat{\xi_{L}}(\sigma) = 0
\end{equation}

where, the constants are given by

\begin{eqnarray}
A_{1} & = & \left (m ^2 + \eta_{F}^2 - \frac{b^2}{4} \right )\\
A_{2} & = & \left (\eta_{F} - \frac{b}{2} \right )^2
\end{eqnarray}

The exact solutions for the massive modes corresponding to the above equation can be 
written in terms of the Hypergeometric functions in the following way

\begin{equation}
\hat\xi_{L} (\sigma) = \mbox{sech} ^{(\frac{\eta_{F}}{2 b} - \frac{1}{4})} (b \sigma) 
 {}^2F_{1} \left [\epsilon - s, \epsilon+s+1, \epsilon + 1, \frac{1}{2} 
(1 - \tanh (b \sigma))\right] 
\end{equation}

where,
\begin{eqnarray}
 \epsilon & = & \left (\frac{\eta_{F}}{b} - \frac{1}{2} \right ) \\
s & = & \frac{1}{2} 
\left (- 1 + \frac{2 \sqrt{m^2 +  \eta_{F}^2}}{b} \right )
\end{eqnarray}

\begin{figure}[htb]
\includegraphics[width= 10cm,height=5.5cm]{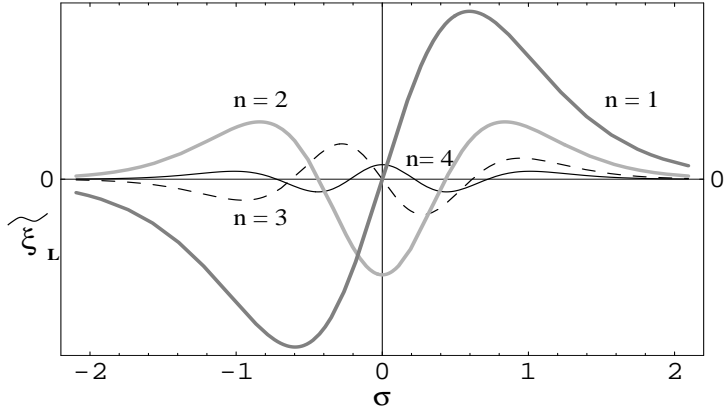}
\caption{The wave function corresponding to the massive left chiral fermions, 
$\tilde\xi_{L}$ is plotted as a function of the extra coordinate $\sigma$ for 
$n = 1, 2, 3$ and $4$ and $\eta_{F} =$ b, 2b, 3b and 4b respectively.} 
\end{figure}

The normalization conditions are obviously satisfied by the massive 
fermions. The modes with mass m will be found confined to the brane till the 
well depth is less compared to the kinetic energy of the states. 
We obtain a mass spectrum from the exact solution obtained in the above case. 
The parity of the chiral fermions with respect to the transformation 
$\sigma \rightarrow - \sigma$ can be obtained from the symmetry property of
the full five dimensional function $\Psi(x^{\mu}, \sigma)$. The full function 
will have a definite parity because of the overall symmetry of the problem 
under $\sigma \rightarrow -\sigma$. Thus the right and left chiral wavefunctions 
will be either odd or even. Mass spectrum of the modes can be obtained from 
condition imposed on the wave function to be well behaved on the brane. 
The possible values of $m^2$ are given by the following expression :

\begin{equation}
m_{n}^2 = \left (n^2 + \frac{2 n \eta_{F}}{b} \right ) b^2
\end{equation}

where n = 1, 2, 3, 4 ..... 

Notice that we have excluded $n = 0$. This is because, the second order equation 
we are solving for is obtained using the restriction $m\neq 0$. For $m=0$,
as mentioned before, the first order equations decouple and have to be 
treated separately. The mass spectrum depends on two parameters $\eta_{F}$ 
and b (or $\vert \Lambda \vert$). The earlier models dealing with the localisation of 
fermions required a large value of coupling constant for the massive modes to be confined 
around the brane. Here, even for $\eta_{F} = 0$, the mass of the fermion is $m_{n} \sim n b$. 
It is evident from the spectrum in Eqn. (31) that heavier masses can be found on the 
brane for stronger coupling.

Let us now summarize pointwise the results obtained in this article and discuss open issues. 

(i) An exact thick brane solution with an increasing warp factor is obtained 
for a bulk scalar field with negative kinetic energy and a 
sine-Gordon potential. 

(ii) Localization of massive and massless fermions are studied 
in this background geometry. The massless left
as well as right chiral modes are found to be normalizable on the brane. 

(iii) Massive modes are also confined to the brane. 
Their mass spectrum is obtained in the 
presence of a Yukawa coupling though they remain 
localizable without the coupling. 

Our article provides a toy model with a growing warp factor for which we
are able to obtain exact analytical results. It is true that growing
warp factors might be problematic in other situations such as the
hierarchy problem, localisation of gravity. In particular, 
one may need to consider the two-brane RS1 set-up to avoid 
embarrasing conclusions related to these problems. 
In such a set-up one may view the brane on which fermions are
localised (via an increasing warp factor) as the `negative tension'
brane with gravity being localised on the `positive tension' brane
situated at a separate location.  
However, at this stage, when one does not even
know clearly whether there is warping in the real world, it is
probably useful to keep in mind the full spectrum of possibilities. 
With this intention, we have, in this article, put forward an exact
increasing warp factor solution and discussed one of its consequences.
We hope to report more results on similar issues in the near future.


\begin{references}

\bibitem{rs} L. Randall and R. Sundrum, Phys. Rev. Lett. {\bf 83}, 3370
(1999); {\em ibid} Phys. Rev. Lett. {\bf 83}, 4690 (1999) ; 


\bibitem{ruba} V. A. Rubakov and M. E. Shaposhnikov, Phys. Lett. {\bf B 125} 136 (1983);
K. Akama, in Proc. Int. Symp. at Nara, Japan, Springer, (1983) 267

\bibitem{increasing} 
M. Visser, Phys. Letts. {\bf B159}, 22 (1985)
M. Maziashvili,hep-th/0404218 ; M. Gogberashvili and
P. Midodashvili, Phys. Letts. {\bf B515}, 447 (2001) {\em ibid.} Europhys.
Letts. {\bf 61},208 (2003) ; P. Midodashvili, hep-th/0308051 ; M. Gogberashvili
and D. Singleton, Phys. Rev. {\bf D69}, 026004 {\em ibid} Phys. Letts. {\bf B
582}, 95 (2004)

\bibitem{cald}  R. R. Caldwell, M. Kamionkowski and N. N. Weinberg, astro-ph/0302506
 
\bibitem{carr} S. M. Carroll, M Hoffman and M. Trodden, astro-ph/0301273; 
G. W. Gibbons, hep-th/0302199; H. P. Nilles, Phys. Rep, {\bf 110}, 1 (1984);
M. D. Pollock,  Phys. Lett. {\bf B 215}, 635 (1988); 
P. Frampton, astro-ph/0209037; 
V. Sahni and Y. Shtanov, astr-ph/0202346; 
B. McInnes, JHEP {\bf 0208}, 029 (2002)

\bibitem{bajc} B. Bajc and G. Gababdadze, Phys. Letts. {\bf 474}, 282 (2000)
\bibitem{rksk} R. Koley and S. Kar, {\em Scalar kinks and fermion localisation
in warped spacetimes} (submitted).

\bibitem{visser} For a useful summary on the details of the energy conditions 
see {\em Lorentzian Wormholes : from Einstein to Hawking} by Matt Visser (AIP, 1995)

\bibitem{rebbi} R. Jackiw and C. Rebbi, Phys. Rev. {\bf D13}, 3398 (1976); Y. Grossman and 
N. Neubert, Phys. Lett. {\bf B474} 361 (2000)

\bibitem{wb} S. Weinberg, {\em
Gravitation and Cosmology} (John Wiley and Sons, 1971)

\bibitem{ringeval} C. Ringeval, P. Peter, J. P. Uzan, Phys. Rev.
{\bf D 65}, 044416 (2002); S. Ichinose, Phys.Rev. {\bf D66}, 104015 (2002)

\bibitem{daemi} S. Randjbar--Daemi and M. Shaposhnikov, Phys. Letts.
{\bf B492}, 361 (2000)

\bibitem{Landau} L.D. Landau and E.M. Lifshitz, {\em{Quantum Mechanics}},
  Course of Theoretical Physics, Vol. 3, Third Edition (Butterworth -
  Heinemann) 
\end{references}
\end{document}